\documentclass[prl,showpacs,floatfix,amsmath,amssymb,twocolumn]{revtex4}
\usepackage{graphicx}
\usepackage{dcolumn}
\usepackage{bm}
\usepackage{amssymb}
\usepackage{float}

\begin{document}

\title{A microscopic view of accelerated dynamics in deformed polymer glasses}

\author{Mya Warren} 
\author{J{\"o}rg Rottler}
\email{jrottler@phas.ubc.ca}

\affiliation{Department of Physics and Astronomy, The University of
  British Columbia, 6224 Agricultural Road, Vancouver, BC, V6T 1Z1,
  Canada}

\begin{abstract} 
A molecular level analysis of segmental trajectories obtained from
molecular dynamics simulations is used to obtain the full relaxation
time spectrum in aging polymer glasses subject to three different
deformation protocols. As in experiments, dynamics can be
accelerated by several orders of magnitude, and a narrowing of the
distribution of relaxation times during creep is directly
observed. Additionally, the acceleration factor describing the 
transformation of the relaxation time distributions is computed and 
found to obey a universal dependence on the global strain, independent 
of age and deformation protocol.
\end{abstract}

\date{\today}
\maketitle 

When amorphous polymers are being deformed, the slow glassy dynamics
resulting from broad distributions of relaxation times becomes
accelerated and permits plastic flow
\cite{gleason2000,yee1988}. Recent experimental advances have made it
possible to directly measure segmental relaxation dynamics in
poly(methyl methacrylate) through an optical photobleaching technique
that probes the reorientation of small dye molecules
\cite{lee2008,ediger2009}. During deformation under applied load,
molecular mobility increases by up to three orders of magnitude with a
concomitant narrowing of the relaxation time spectrum
\cite{lee2008,ediger2009}. Additionally, shifts in mobility relative
to the undeformed glass strongly correlate with the instantaneous
strain rate. In order to explain plasticity in amorphous solids, the
literature frequently invokes Eyring's phenomenological picture of
stress-biased activated processes, in which the viscosity and hence
relaxation times $\tau_r$ decrease with applied stress $\sigma$ as
$\tau_r \sim \sigma/\sinh{[\sigma V/k_BT]}$ \cite{eyring1936}. It has
been demonstrated multiple times, however, that this simple model
fails to predict the yield behavior of polymer glasses consistently
over a range of stresses, strain rates and temperatures
\cite{schweizer_epl07}. For instance, relaxation times change even
when the applied stress is constant \cite{ediger2009}.

Enhanced mobility in deformed polymer glasses has also been observed
in both atomistic and coarse grained computer simulations. Chain
dynamics was probed by measuring the monomer mean-squared displacement
\cite{lyulin2005}, the dihedral transition rate between trans/gauche
states \cite{capaldi2002}, and the decay rate of covalent bond
autocorrelations and intermediate scattering functions
\cite{riggleman2007,riggleman2008}. These very different observables
all report increased local mobility on the intra- and interchain level
both in the sub-yield \cite{capaldi2002} and post-yield
\cite{lyulin2005} regime and confirm the experimentally observed
strong correlation between relaxation time and strain rate
\cite{riggleman2007,riggleman2008}. A limitation of these studies is,
however, that they only report spatially averaged quantities and do
not reveal the full spectrum of relaxation times.  Furthermore, since
even undeformed glasses are not in equilibrium, it is important to
consider the interaction between deformation and physical aging. Aging
increases the stiffness of glassy polymers by increasing structural
relaxation times; however, deformation alters the intrinsic aging
dynamics, and plastic flow can erase the history of the glass, an
effect known as ``mechanical rejuvenation''
\cite{struik1978,lyulin2007}.

In the present Letter, we use molecular simulations as a computational
microscope to obtain detailed insight into the dynamics of polymer
glasses under deformation, and its consequences for physical
aging. From an analysis of individual particle trajectories, we
identify relaxation events and hence obtain the full probability
distribution of relaxation times. We consider three different
deformation protocols: a step stress (creep), a constant strain rate
deformation, and a step strain (stress relaxation).  The relaxation
time distributions are changed in complex ways which depend on the
wait time and the deformation protocol, however we find that these changes
can all be accounted for using a universal acceleration factor which
depends only on the global strain.

\begin{figure*}[t] 
\begin{centering} 
\includegraphics[width=17cm]{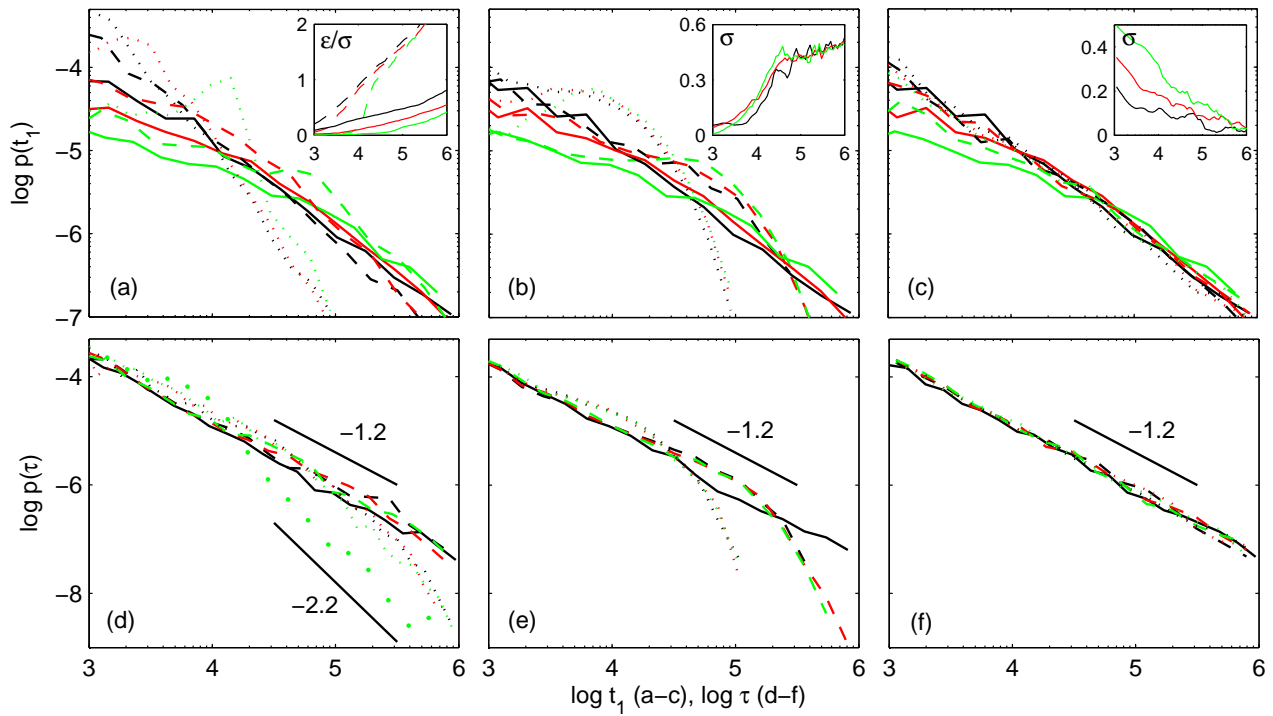}
	\caption{\label{dist-fig}(Color online) Distributions of first
          hop times $t_1$ (a)-(c) and persistence times $\tau$ (d)-(f)
          in an aging polymer glass subject to deformation after
          undergoing physical aging for three waiting times $t_w=750$
          (black), 7500 (red/gray), and 75000 (green/light gray).
          Panels (a) and (d) correspond to a step stress with
          amplitude $\sigma=0.4$ (dashed) and $0.5$ (dotted).  Solid
          dots show $p(\tau)$ for $t<\tau_\alpha$ (see text) and $t_w
          = 75000$.  Panels (b) and (e) refer to deformation at
          imposed strain rate $\dot\epsilon= 8.9\times 10^{-7}$
          (dashed) and $8.9\times 10^{-6}$ (dotted), while panels (c)
          and (f) show the effect of a step strain with amplitude
          $\epsilon=0.02$ (dashed) and $0.04$ (dotted) For comparison,
          the corresponding distributions in an undeformed glass that
          is only undergoing physical aging are also shown in every
          panel (solid lines). Straight lines indicate power law with
          the given slopes. Insets show the mechanical response to (a)
          $\sigma = 0.4$ (solid) and 0.5 (dashed), (b) $\dot\epsilon=
          8.9\times 10^{-7}$, and (c) $\epsilon=0.04$ for the three
          wait times. }
\end{centering} 
\end{figure*}

Our study employs a well-known coarse-grained bead-spring polymer
system, which has been used extensively to model glassy dynamics
\cite{baschnagel05} and polymer deformation \cite{hoy2006}. The beads
are bonded via a finite extensible non-linear elastic (FENE) spring
and interact via a truncated 6-12 Lennard Jones (LJ) potential $V_{\rm
  LJ}(r) = 4u_0 \left[(a/r)^{12} -(a/r)^{6}\right]$ for $r <r_c=1.5$,
where $u_0$ and $a$ set the reference energy and length scale. Results
will be given in LJ units, where the reference time scale is
$\tau_{LJ} = \sqrt{ma^2/u_0 }$. Individual samples consist of 20,000
monomers forming short chains of length 10 in a cubic box subject to
periodic boundary conditions.  After equilibrating at a melt
temperature $T=1.2$, the glass is prepared through a rapid quench at
constant volume to a glassy temperature of $T=0.25$, where the
hydrostatic pressure is close to zero. The glass is then aged at zero
pressure for a waiting time $t_w$ before a uniaxial deformation is
applied in the form of {\it (i)} a step stress of magnitude $\sigma$,
{\it (ii)} a constant strain rate $\dot\epsilon$, and {\it (iii)} a
step strain of magnitude $\epsilon$. In all three protocols the stress
components perpendicular to the deformation axis are kept at zero,
while stress/strain/strain rate are maintained at the indicated
amplitudes.

In amorphous solids, the motion of individual particles is not smooth,
but highly intermittent. Atoms spend long times in caged environments
before undergoing a rapid structural relaxation to a new
position. These relaxations correspond to hops on a segmental
trajectory. We have developed an algorithm that reliably identifies
such hops through peaks in the standard deviation of the particle
position (see ref.~\cite{mya2009} for details). The analysis yields the
full temporal sequence of relaxation events for a subset of 5000
polymer beads chosen randomly from the full simulation volume, from
which relaxation time distributions are readily constructed.

In Fig.~\ref{dist-fig} we show the observed distributions of hop times
for the three deformation protocols defined above and the
undeformed glass. We separately present the distribution of first hop
times $p(t_1)$ since the onset of deformation (panels (a)-(c)) and the
distribution of times in between all subsequent hops $p(\tau)$ (panels
(d)-(f)). $\tau$ measures the lifetime of an individual caged
configuration, which we call the persistence time. The mechanical
response of each protocol is shown as an inset in panels (a)-(c). In
ref.~\cite{mya2009}, we showed that for simple aging with no
deformation, the first hop times increase with a power law in the wait
time, $t_1 \sim t_w^\mu$, where $\mu$ is the aging exponent. The
persistence time distribution, however, is age independent and takes
the form of a power law $p(\tau)\sim \tau^{-1.2}$. Since the mean hop
time of such a distribution is infinite, the dynamics is
non-stationary and ages. Studying the changes to these distributions
under deformation provides rich insight into the origin of
deformation-accelerated dynamics.

First, we consider the effects of a constant stress deformation. The
inset of Fig.~\ref{dist-fig}(a) shows that the strain increases with a
characteristic timescale $\tau_\alpha$, which grows with wait time as
$t_w^\mu$. In the sub-yield regime, the response remains shifted in
time for different wait times, whereas above yield, the glass
eventually begins to flow and all aging is erased (mechanical
rejuvenation).  The first hop time distribution $p(t_1)$
(Fig.~\ref{dist-fig}(a)) is significantly narrowed for increasing
stress. Not only is the probability to observe short first hop times
increased relative to the undeformed sample, but the power law tail 
decays more rapidly. In this way, our analysis permits direct
observation of the narrowing of the relaxation time spectrum during
creep deformation, which previously was inferred only indirectly from
stretched exponential fits to correlation functions \cite{ediger2009}.

Fig.~\ref{dist-fig}(d) shows the persistence time distributions, which
for a constant stress deformation depend explicitly on both wait time
and the time since the deformation began (measurement time $t$).
Lines in Fig.~\ref{dist-fig}(d) show $p(\tau)$ for cages that come
into being after the alpha relaxation time $(t>\tau_\alpha)$, when the
strain is increasing slowly. Similar to the first hop time
distribution, $p(\tau)$ is narrowed and decreases faster for large
$\tau$. This effect is even more pronounced when $p(\tau)$ is
calculated from particles whose cages were created at $t<\tau_\alpha$
(dots).  The narrowing of the persistence time distributions provides
an interesting explanation for mechanical rejuvenation.
Our results show that at $\sigma = 0.5$, where significant plastic
flow occurs, the tails of $p(\tau)$ decay with an exponent of $\sim
-2.2$. For such a distribution, the mean persistence time becomes
finite and a steady state is reached which does not depend on the
history of the sample.

\begin{figure}[t] 
\begin{centering} 
\includegraphics[width=8cm]{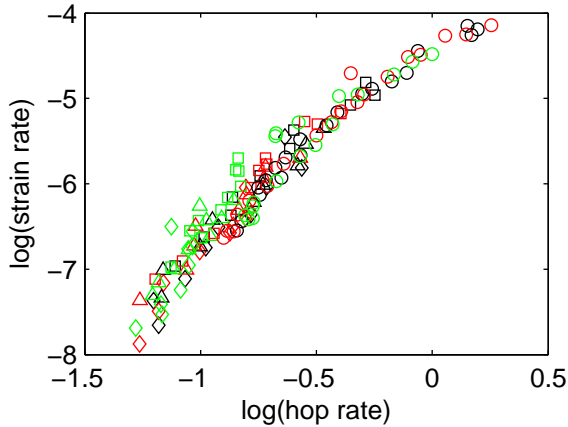}
	\caption{\label{hop_vs_strain_rate-fig} (Color online) Strain
          rate versus hop rate for $\sigma = 0.5$ ($\circ$), 0.4
          ($\square$), 0.3 ($\triangle$), 0.2 ($\lozenge$) and $t_w =
          750$ (black), 7500 (red/gray), and 75000 (green/light
          gray). }
\end{centering} 
\end{figure}

Understanding the dynamics during creep has been the focus of several
recent investigations \cite{lee2008,ediger2009,riggleman2008}. In
these works, the mobility is defined as the inverse of the average
relaxation time, computed through repeated measurements of an
autocorrelation function during the course of the experiment. Results
show that the mobility increases almost linearly with the strain rate,
independent of stress and temperature. The mobility as defined in
these experiments is conceptually similar to the total hop rate,
accessible through our technique. In
Fig.~\ref{hop_vs_strain_rate-fig}, the strain rate is plotted versus
the hop rate during the step stress experiment.  Results for all
stresses and wait times fall on the same, universal curve.
The curve is not exactly linear due to the presence of a finite hop
rate even in the undeformed, aging samples, and an increase of the
mean (non-affine) displacement per hop with strain rate.
 

\begin{figure}[t] 
\begin{centering} 
\includegraphics[width=8cm]{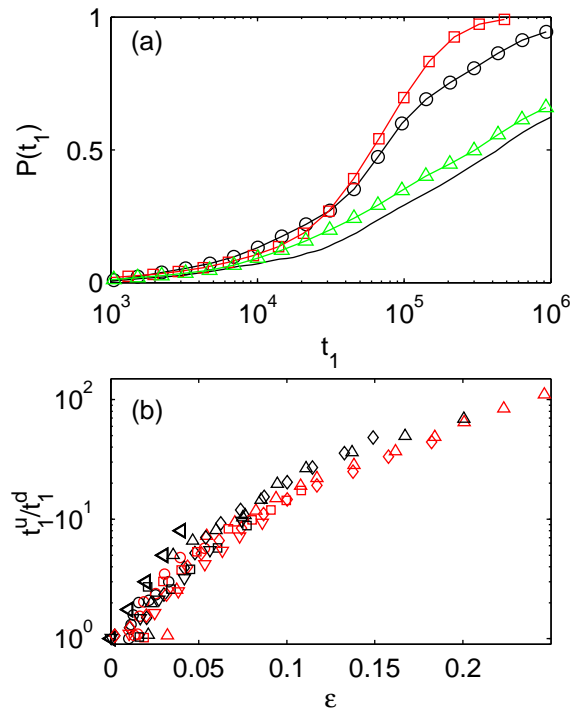}
	\caption{\label{accel-fig}(Color online) (a) Cumulative
          probability distribution of the first relaxation event for
          the undeformed glass (solid line) and under deformation with
          a step stress $\sigma=0.4$ $(\circ)$, a step strain
          $\epsilon=0.01$ $(\triangle)$, and a constant strain rate
          $\dot\epsilon=8.5\times 10^{-7}$ $(\square)$ for
          $t_w=75000$. (b) Acceleration ratio $t_u/t_d$ as a function
          of global strain $\epsilon$ for three different deformation
          protocols. Stress step: $\sigma = 0.3$ ($\circ$), 0.4
          ($\square$), 0.5 ($\triangle$); constant strain rate:
          $\dot{\epsilon} = 8.9\times 10^{-6}$ ($\lozenge$), $8.9
          \times 10^{-7}$ ($\triangledown$); strain step
          ($\triangleleft$).  For each: $t_w = 75000$ (black), 22500
          (red).}
\end{centering} 
\end{figure}

A constant strain rate deformation produces a very different effect on
the hop times than the creep experiment. Now the system is undergoing
plastic flow at an externally imposed rate, and this timescale is
reflected in the relaxation time distributions. One can see in
Fig.~\ref{dist-fig}(b) that the tails of the $p(t_1)$-distributions
steepen considerably at times of order $0.1/\dot\epsilon$. At short
times, however, the distributions are unmodified relative to the
unperturbed glass. Similarly, the persistence time distributions
$p(\tau)$ (Fig.~\ref{dist-fig}(e)) are not modified at short times,
but decrease rapidly over the same timescale as $p(t_1)$. $p(\tau)$
remains independent of the wait time and the measurement time. Aging
is also erased in this protocol, but now due to truncation of the
power law distribution rather than a change in exponent as was
observed for constant stress. These results are completely consistent
with the effect of aging on the macroscopic mechanical behavior (inset):
stress-strain curves display an initial overshoot stress
that increases with age, followed by an age independent flow or
hardening regime.

Finally, we analyze the hop statistics for the step strain
experiment. Fig.~\ref{dist-fig}(c) shows that increasing the amplitude
of the strain step primarily modifies the short time part of
$p(t_1)$. Alternatively, the persistence times
(Fig.~\ref{dist-fig}(f)) are completely unchanged relative to the
undeformed glass. The effects of a strain step on the first hop time
can more easily be appreciated through the cumulative distributions
$P(t_1) = \int_0^{t_1} p(t_1^\prime) dt_1^\prime$. $1-P(t)$ is the 
probability that a particle has not yet hopped at time $t$, and its decay
closely resembles the autocorrelation functions more typically used 
to measure particle relaxations such as the intermediate scattering 
factor \cite{mya2010}. Figure \ref{accel-fig}(a) 
shows that the cumulative distribution in the step strain protocol is 
simply shifted in time by a constant factor with respect to the 
undeformed glass: all relaxation times are rescaled identically.

The cumulatives for the step stress and the constant strain rate
protocol are also shown in Figure \ref{accel-fig}(a). Unlike the
simple time shift seen in the step strain experiment, undeformed and
deformed curves continue to diverge with time. To make the
transformation of the relaxation time distributions more quantitative,
we define an acceleration ratio as the ratio of times when the
undeformed and deformed cumulative distributions are equal:
$t_1^u/t_1^d$, where $t_1^u$ and $t_1^d$ are defined by the
relationship $P(t_1^u) = P(t_1^d)$. The acceleration ratio captures
changes to the full distribution of relaxation times and is therefore
distinct from the average mobility defined in earlier works.
Remarkably, the variation in
acceleration ratio for the different deformation protocols can be
reduced to the effect of their global strain. A parametric plot of the
ratios $t_1^u/t_1^d$ versus the total strain experienced by the sample
is shown in Fig.~\ref{accel-fig}(b) for all three protocols and two
different waiting times $t_w$ \cite{note}. Despite the dramatically
different mechanical response, all data sets collapse reasonably well
on a common curve of the form $t_1^u/t_1^d\sim \exp(f(\epsilon))$,
where $f$ is a universal function of strain. We were unable to achieve
a similar collapse using any other deformation variable. For sub-yield
strains, we observe a close to exponential rise in the acceleration
ratio as a function of global strain. The acceleration increases more
slowly after strains of approximately five percent.

This result also explains the observed changes to the persistence time
distributions. In this case, the relevant parameter is the accumulated
strain in between subsequent hops. For instance, a strain step causes
no change to $p(\tau)$ since, after the first hop, there is no further
strain. In the constant strain rate experiment, the acceleration
factor can be defined in the same way as before, using instead the
cumulatives of $p(\tau)$. Plotted versus the strain, this gives the
same universal curve as in Fig.~\ref{accel-fig}(b) demonstrating that
the second hops are accelerated in exactly the same way as the first
hops. Although it is more difficult to quantify, the dependence of
$p(\tau)$ on the wait time and the measurement time in the step stress
experiment can be explained by the fact that the strain explicitly
depends on these same variables.

The universal collapse observed in Fig.~\ref{accel-fig}(b) strongly
suggests that the global strain is a good variable to describe the
influence of deformation on structural relaxations, rather than the
stress as postulated by the Eyring model. It is interesting to note
that the strain was chosen as control variable in the Soft Glassy
Rheology (SGR) model \cite{sollich1997}. Here the acceleration factor
for barrier crossings increases proportional to $\exp(l^2)$, where
$l$ measures the local strain on a mesoscopic element. In
this model, the local strains increase in tandem with the global
strain in between barrier crossings. Although the measured
acceleration factor differs from that of the SGR model, inspection of
our particle trajectories shows that this may be a good picture of
local strain in our system: individual particles exhibit very little
non-affine displacement except during a hop.

We have presented the first investigation of the full relaxation time
distribution of aging, deformed polymer glasses. As reported
previously, all three deformation modes accelerate the segmental
dynamics. Although the exact transformation of the relaxation time
distributions is specific to the mode of deformation, we suggest that
the acceleration can be described by a universal function of the
global strain. Control simulations with longer chains containing 100
monomers indicate that this conclusion holds independent of chain
length \cite{mya2010}. We also provide a possible explanation for the
phenomenon of mechanical rejuvenation, whereby aging is arrested by
deformation.  The persistence time distribution causes aging in
glasses because of the weak power law decay. Ergodicity is restored
during flow under both a step stress and constant strain rate
deformation because this distribution is narrowed and the mean
persistence time becomes finite.

This work was supported by the Natural Sciences and Engineering
Research Council of Canada (NSERC).


\begin{thebibliography}{19}
\bibitem[{not()}]{note}
\bibinfo{note}{The acceleration factor can be accurately determined only when
  hop data is well resolved at times well below the cage escape time. For this
  reason, acceleration factors for only the longest $t_w$ are shown.}
\end{thebibliography}

\end{document}